\documentclass[conference]{IEEEtran}

\usepackage[acronym]{glossaries}
\usepackage{cite}
\usepackage{siunitx}
\usepackage{graphicx}
\usepackage[caption=false,font=footnotesize]{subfig}
\usepackage{xcolor}
\usepackage{booktabs}
\usepackage{multirow}
\usepackage{tikz}
\usepackage{comment}
\usepackage{pifont}
\usepackage{stfloats}
\usepackage{balance}

\newacronym{ai}{AI}{Artificial Intelligence}
\newacronym{ml}{ML}{Machine Learning}
\newacronym{eeg}{EEG}{ElectroEncephaloGram}
\newacronym{ecg}{ECG}{ElectroCardioGram}
\newacronym{ppg}{PPG}{PhotoPlethysmoGraphy}
\newacronym{imu}{IMU}{Inertial Measurement Unit}
\newacronym{emg}{EMG}{ElectroMyoGraphy}
\newacronym{pcb}{PCB}{Printed Circuit Board}
\newacronym{i2c}{I2C}{Inter-Integrated Circuit}
\newacronym{d2m}{D2M}{Daughter-to-Main}
\newacronym{d2d}{D2D}{Daughter-to-Daughter}
\newacronym{de2m}{DE2M}{Debug-to-Main}
\newacronym{w2b}{W2B}{Wire-to-Board}
\newacronym{uart}{UART}{Universal Asynchronous Receiver-Transmitter}
\newacronym{mcu}{MCU}{Main Computing Unit}
\newacronym{ble}{BLE}{Bluetooth Low Energy}
\newacronym{led}{LED}{Light-Emitting Diode}
\newacronym{rtos}{RTOS}{Real-Time Operating System}
\newacronym{spi}{SPI}{Serial Peripheral Interface}
\newacronym{iot}{IoT}{Internet of Things}
\newacronym{mpu}{MPU}{Motion Processing Unit}
\newacronym{hr}{HR}{Heart Rate}
\newacronym{ir}{IR}{Infrared}
\newacronym{spo2}{SPO2}{Peripheral Oxygen Saturation}
\newacronym{rmse}{RMSE}{Root Mean Squared Error}
\newacronym{gpio}{GPIO}{General-Purpose Input/Output}
\definecolor{sensgray}{HTML}{CCCCCC}
\newcommand{\circled}[3][2.2ex]{%
  \tikz[baseline=(C.base)] \node[draw,circle,inner sep=0pt,
    minimum size=#1, line width=#3] (C) {\strut #2};%
}

\newcommand{\blackcircle}[2][2.4ex]{%
  \tikz[baseline=(C.base)] 
    \node[draw,
          circle,
          fill=black,
          text=white,
          inner sep=0pt,
          minimum size=#1,
          line width=0.6pt] (C) {\strut #2};%
}

\newcommand{\graycircle}[2][2.4ex]{%
  \tikz[baseline=(C.base)] 
    \node[
      circle,
      draw=black,
      fill=sensgray,
      text=black,
      inner sep=0pt,
      minimum size=#1,
      line width=1.2pt
    ] (C) {\strut #2};%
}
\providecommand{\checkmark}{}
\renewcommand{\checkmark}{\ding{51}}
\AtBeginDocument{\renewcommand{\checkmark}{\ding{51}}}

\usepackage{xspace}
\newcommand{\platformname}{SensWear\xspace}

\begin{document}
%
\title{\platformname: An Open, Modular, and AI-Ready \\Wearable Platform}
%
%
%

\author{Dariush~Salami,
Behzad~Salami,~and~H{\"u}seyin~Yi{\u{g}}itler
}

\glsresetall
\maketitle

\begin{abstract}
Wearable research requires raw and synchronized multimodal data across diverse wearable form factors. However, the wearable ecosystem remains fragmented across fixed embodiments, sensor configurations, and software stacks, while commercial devices further restrict access to hardware, firmware, and raw data. This paper presents \textit{\platformname}, an open, modular, and AI-ready wearable platform that aims to unify fragmented research workflows by decoupling embodiment, sensing, data interfaces, and computation. As part of this work, we designed and implemented the complete \platformname hardware, including a flexible-PCB main board based on low-power nRF54L15 SoC with an onboard \gls{imu} and multiple flexible-PCB daughter boards for \gls{ppg}, touch, body temperature, and haptic feedback, interconnected through a common I2C bus. Together with a unified firmware and software stack, this architecture enables different wearable form factors and custom sensor configurations on a common platform. To demonstrate one possible embodiment, we implement \platformname as a custom ring and comprehensively characterize its power consumption, data streaming, local storage, and on-device AI performance. The platform consumes 0.436~mA when connected and idle and 3.153~mA during simultaneous \gls{imu} and \gls{ppg} streaming. Both \gls{imu} and \gls{ppg} streaming achieve 100\% measured data completeness in the evaluated trials, while a representative CNN executes in 8.06~ms, corresponding to 124.1 inference windows/s. End-to-end case studies further demonstrate raw arterial \gls{ppg} acquisition and resting heart-rate estimation. The complete hardware, firmware, and software stack is open source under the MIT License.
\end{abstract}
\glsresetall
\section{Introduction}
\label{sec:introduction}

\begin{figure}[!t]
    \centering
    \includegraphics[width=0.8\linewidth]{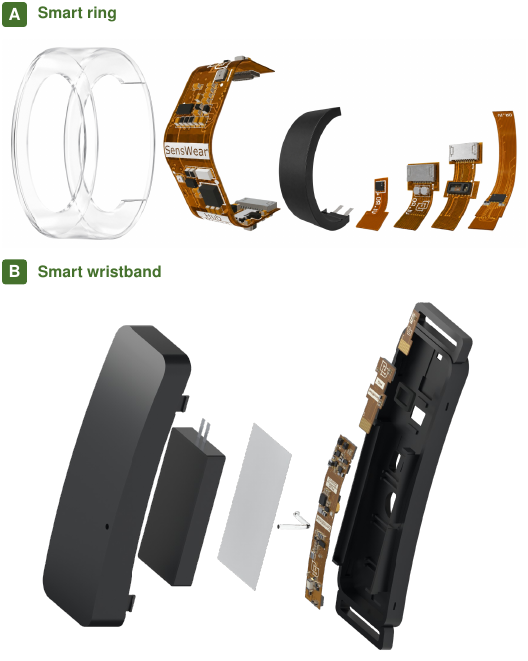}
    \caption{\platformname modular wearable platform shown in two embodiments. (A) Smart-ring view showing the enclosure, flexible main board, curved battery, and daughter-board alternatives for body temperature, haptic feedback, \gls{ppg}, and touch, from left to right. (B) Smart-wristband view showing the enclosure, battery, main board, and \gls{ppg} and body temperature daughter boards.}
    \vspace{-5 mm}
    \label{fig:platform_teaser}
\end{figure}

Wearable sensing is central to human-centric computing, health, and ubiquitous interaction, and recent \gls{ai} methods increasingly require raw, synchronized, multimodal data across diverse wearable form factors and real-world conditions~\cite{bahmani2025wearable,salami2022tesla}. However, the wearable ecosystem remains fragmented across fixed embodiments, sensor configurations, and software stacks. Commercial devices further restrict access to hardware, firmware, and raw data, while many research platforms are designed around specific form factors or sensing modalities~\cite{konig2023let,zhou2023one,roddiger2022openearable}. Consequently, exploring different sensing modalities, body placements, or wearable form factors often requires adopting or redesigning different platforms, fragmenting research workflows and limiting reproducibility and reuse.

This paper presents \textit{\platformname}, an open, modular, and AI-ready wearable platform that aims to unify fragmented wearable research workflows through a common hardware and software architecture. \platformname is designed as reusable infrastructure in which form factor, sensing, and computation can be independently configured. Its compact, low-power, flexible-PCB main board integrates the nRF54L15 SoC, an onboard \gls{imu} for motion sensing, and local storage. Onboard RGB feedback and daughter-board haptics further enable bidirectional interaction. Sensor modularity is provided through interchangeable flexible-PCB daughter boards sharing a common I2C bus, GPIO, and programmable power, with current boards supporting \gls{ppg}, capacitive touch, body temperature sensing, and haptic feedback, while allowing additional sensor modules to be integrated. Together, the conformable flexible-PCB design and disaggregated architecture enable \platformname to be adapted to multiple wearable form factors, including ring, wrist, patch, and custom embodiments, with configurable sensor sets. A unified Zephyr-based firmware and software stack, including Python and JavaScript/TypeScript SDKs, provides raw synchronized data, timestamping, local storage, and support for edge, mobile, or cloud processing. These capabilities provide an end-to-end path from configurable multimodal sensing to \gls{ai} processing and application development, enabling applications across different domains while retaining common hardware and software interfaces. The complete hardware, firmware, and software stack is released under the MIT License to support openness, reproducibility, and further customization. Figure~\ref{fig:platform_teaser} illustrates this flexibility through smart-ring and smart-wristband embodiments of \platformname. The contributions are:

\begin{itemize}
    \item A customizable wearable platform with fully open-source HW/SW stack.

    \item A modular system design that enables independent evolution of the wearable, sensing, and computing layers.

    \item A fully implemented and experimentally validated platform, demonstrated through a functional ring embodiment (as an demonstration example), with characterization of functionality, performance, and power consumption as well as end-to-end application.
\end{itemize}


\begin{figure*}[!t]
    \centering

    \includegraphics[width=\linewidth]{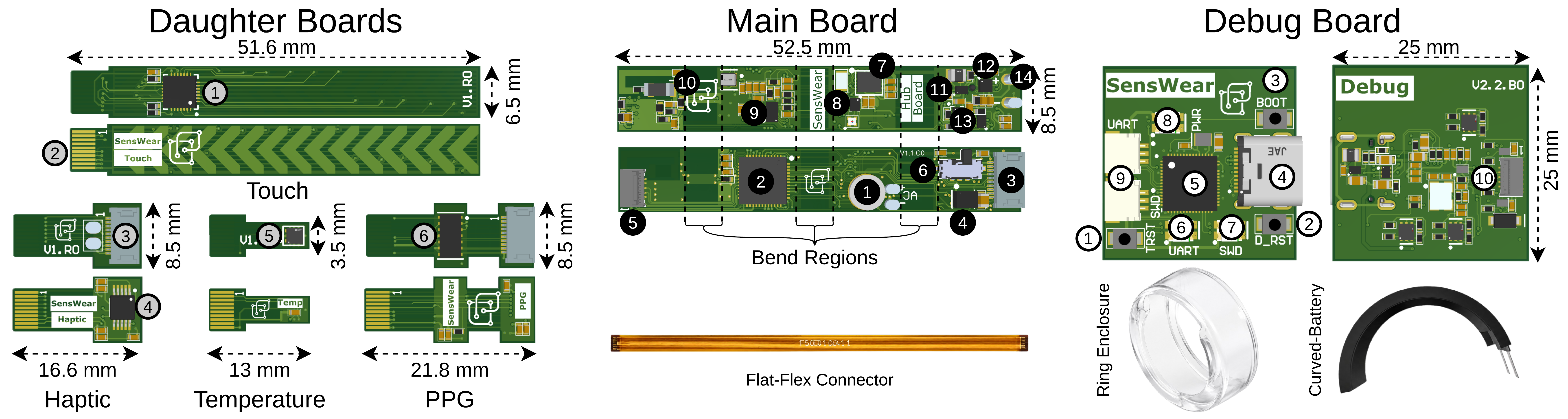}
    \caption{Overview of the wearable platform including the main board, daughter boards, debug board,c onnector, ring enclosure, and curved battery.}
    \label{fig:system_model}
    \vspace{-5 mm}
\end{figure*}

\begin{table}[!t]
    \centering
    \caption{Platform components and interfaces.}
    \begin{tabular}{@{}llll@{}}
    \toprule
        & ID & Component & Description \\ 
    \midrule
    \multirow{12}{*}{\rotatebox[origin=c]{90}{Main}}
        & \blackcircle[2.4ex]{1}  & Pogo             & Pogo connector for charging \\
        & \blackcircle[2.4ex]{2}  & NRF54l15         & Wireless SoC and MCU \\
        & \blackcircle[2.4ex]{3}  & D2M*             & F. connector for daughter boards \\
        & \blackcircle[2.4ex]{4}  & TPSM831021       & Voltage regulator \\
        & \blackcircle[2.4ex]{5}  & DE2M*            & F. connector for the debug board \\
        & \blackcircle[2.4ex]{6}  & SSAJ110100       & Board switch \\
        & \blackcircle[2.4ex]{7}  & BHI360           & 6D-\gls{imu} for motion processing \\
        & \blackcircle[2.4ex]{8}  & LP5562           & LED driver \\
        & \blackcircle[2.4ex]{9}  & M95P32-IXCST     & 32 Mbit EEPROM memory \\
        & \blackcircle[2.4ex]{10}  & 2450AT18E0100E   & 2.4GHz antenna for BLE \\
        & \blackcircle[2.4ex]{11} & TPS62840         & HE step-down converter \\
        & \blackcircle[2.4ex]{12} & BQ27427          & Battery fuel gauge \\
        & \blackcircle[2.4ex]{13} & BQ25180          & Linear battery charger \\
        & \blackcircle[2.4ex]{14} & Pads          & Battery pads \\
    \midrule
    
    \multirow{6}{*}{\rotatebox[origin=c]{90}{Daughters}}
        & \graycircle{1} & MTCH6102 & Touch controller \\
        & \graycircle{2} & D2M*     & M. connector \\
        & \graycircle{3} & D2D*     & F. connector \\
        & \graycircle{4} & DRV2605  & Haptic driver \\
        & \graycircle{5} & MAX30208 & Body temperature sensor \\
        & \graycircle{6} & MAX30101 & \gls{ppg} sensor \\
    \midrule

    \multirow{5}{*}{\rotatebox[origin=c]{90}{Debug}}
        & \circled[2.4ex]{1}{1.2pt} & B3U-1000P    & Reset for the target \\
        & \circled[2.4ex]{2}{1.2pt} & B3U-1000P    & Reset for the debug board \\
        & \circled[2.4ex]{3}{1.2pt} & B3U-1000P  & Boot mode selection \\
        & \circled[2.4ex]{4}{1.2pt} & USB  & USB connector \\
        & \circled[2.4ex]{5}{1.2pt} & RP2354A & Debugger MCU \\
        & \circled[2.4ex]{6}{1.2pt} & C155KGJRKT & LED for UART status \\
        & \circled[2.4ex]{7}{1.2pt} & C155KGJRKT & LED for SWD status \\
        & \circled[2.4ex]{8}{1.2pt} & C155KGJRKT & LED for power status \\
        & \circled[2.4ex]{9}{1.2pt} & SM03B-SRSS-TB & Headers for UART and SWD \\
        & \circled[2.4ex]{10}{1.2pt} & Connector & Debug board to main board \\
    \midrule
    \addlinespace[2pt]
    \multicolumn{4}{@{}p{\columnwidth}@{}}{%
        \footnotesize
        D2M: Daughter-to-Main, DE2M: Debug-to-Main, D2D: Daughter-to-Daughter, F: Female, M: Male, HE: High-Efficiency.
        
    } \\
    \bottomrule
    \vspace{-8 mm}
    \label{tab:components_table}
    \end{tabular}
\end{table}

\section{\platformname System Overview}
\label{sec:system_overview}
Figure~\ref{fig:system_model} shows the hardware components of \platformname: a main board for processing, wireless communication, memory, and power management; interchangeable sensing and actuation daughter boards; and a debug board for programming, reset, and UART, flat-flex connector for connecting the main boad to the daughter board, a smart-ring enclosure, and a 23mAh curved battery. Table~\ref{tab:components_table} identifies the components and the \gls{d2m}, \gls{d2d}, \gls{de2m}, and \gls{w2b} interfaces.

\subsection{Hardware Architecture}

\begin{figure*}[t]
    \centering
    \includegraphics[width=0.9\linewidth]{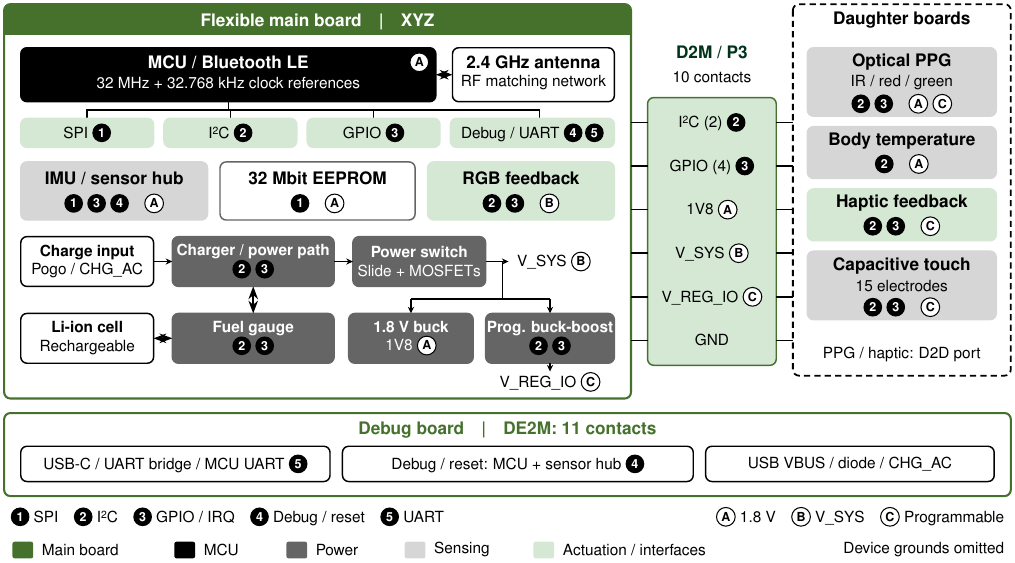}
    \caption{\platformname hardware architecture, showing the main board, interchangeable daughter boards, and debug board.}
    \label{fig:hardware_architecture}
    \vspace{-5 mm}
\end{figure*}

Figure~\ref{fig:hardware_architecture} separates the main board's shared computing, power, and motion-sensing functions from daughter boards. The \textit{nRF54L15} integrates the low-power \gls{mcu} and \gls{ble} radio. The \textit{BHI360} provides motion sensing and embedded fusion, while the \textit{M95P32} provides \qty{32}{\mega\bit} of EEPROM storage; both share an \gls{spi} bus with separate chip selects. A common \gls{i2c} bus connects power-management devices, the LP5562 driver for RGB feedback, and daughter-board peripherals.

In Figure~\ref{fig:hardware_architecture}, the rechargeable cell connects to the BQ27427 fuel-gauge interface, whose system-side connection leads to the BQ25180 charger/power-path device. A slide switch and MOSFETs gate the system supply. The TPS62840 produces a fixed \qty{1.8}{\volt} logic rail, and the TPSM831021 produces the programmable daughter-board rail. The ten-contact \gls{d2m} connector carries these two rails, the system supply, \gls{i2c}, four \gls{gpio} lines, and ground.

Daughter boards include MAX30101 \gls{ppg}, MAX30208 body temperature, MTCH6102 capacitive touch, or DRV2605L haptic feedback. The \gls{ppg} and haptic boards expose \gls{d2d} extension connectors; shared supply and \gls{gpio} assignments determine which combinations can operate together. This interface lets researchers change sensing or feedback modules while retaining the main board. The flexible-PCB footprint and bend regions support custom enclosures. The debug board connects through the eleven-contact \gls{de2m} interface and provides UART for debugging purposes, external debug headers for the MCU and BHI360, reset buttons, and charging input to support development without battery or additional power cables.

For a custom wearable, the same main board is reused with a selected set of existing or newly designed daughter boards connected through the common D2M interface. The daughter boards can be supplied through the programmable daughter-board rail, and share the common I²C bus, and in some cases they can be chained through D2D connections.

\subsection{\platformname Firmware, Data, and \gls{ai} Interfaces}
\label{sec:edge_ai}

\begin{figure*}[t]
    \centering
    \includegraphics[width=0.9\linewidth]{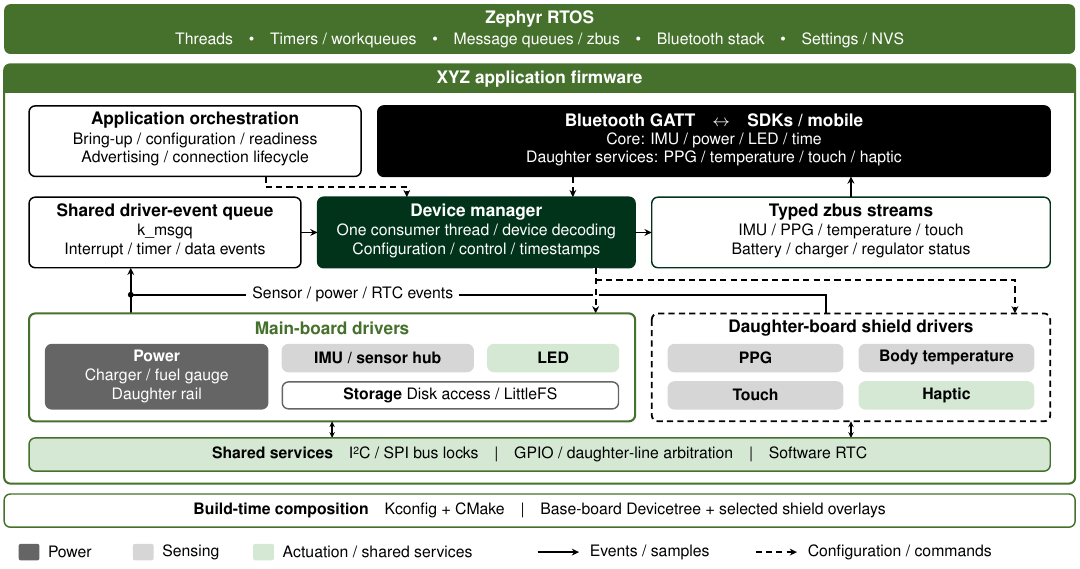}
    \caption{\platformname firmware architecture.}
    \label{fig:firmware_architecture}
    \vspace{-5 mm}
\end{figure*}

Figure~\ref{fig:firmware_architecture} follows acquisition and control within a single Zephyr \gls{rtos} application on the \gls{mcu}. Sensor, power, and timer events enter a shared driver-event queue. A device-manager thread handles the deferred work and publishes decoded, timestamped messages on typed zbus channels, which the \gls{ble} GATT services consume. Configuration, clock updates, and actuation follow the path shown by dashed arrows through device-manager APIs.

The lower layers in Figure~\ref{fig:firmware_architecture} explain how this data path accommodates different daughter boards. Main-board and shield drivers use shared bus locks, connector-line arbitration, and software timekeeping. Kconfig, CMake, and Devicetree overlays select shield drivers and corresponding services at build time. The \gls{ppg} preset enables both MAX30101 and MAX30208 (body temperature) shields for a combined configuration using separate \gls{ppg} and body temperature boards. The software real-time clock supplies a common timestamp reference, while the M95P disk-access adapter exposes LittleFS file system for application file access independently of sensor streaming.

The same typed streams provide inputs for application-specific feature extraction and inference on the \gls{mcu}, smartphone, or cloud, alongside the BHI360's embedded motion processing. Separating acquisition, transport, and application analysis allows researchers to change an algorithm's execution location while retaining the sensor configuration and data interface. Local file access provides infrastructure for application-defined recording; automatic recording of every stream is not part of the acquisition path shown in the figure.

For end-to-end \gls{ai} applications, \platformname exposes the same timestamped multimodal sensor streams across multiple execution locations. Lightweight feature extraction and inference can execute locally on the nRF54L15, while raw or processed streams are exposed through BLE and the SDK for more computationally intensive processing on mobile, edge, or cloud systems. The SDKs provide access to sensor configuration and data streams, enabling data collection, model development, inference, and application integration without changing the underlying sensing interface. This decoupling allows the sensing configuration and \gls{ai} processing location to be selected independently for different application requirements.

\section{Experimental Analysis on \platformname}
\label{sec:performance}

This section experimentally evaluates \platformname using a fully implemented ring as one example wearable form factor. The goal is twofold: to validate the complete end-to-end functionality of the platform and to quantify its key operating characteristics and resource budgets. We characterize the platform with four complementary methods: power consumption under different workloads, end-to-end data streaming performance, local-storage performance, and on-device \gls{ai} computation. Battery-life estimates are derived from the measured mean current and the nominal \qty{23}{\milli\ampere\hour} battery capacity.

\subsection{Operational Current}

\begin{figure}[!t]
    \centering
    \includegraphics[width=\linewidth]{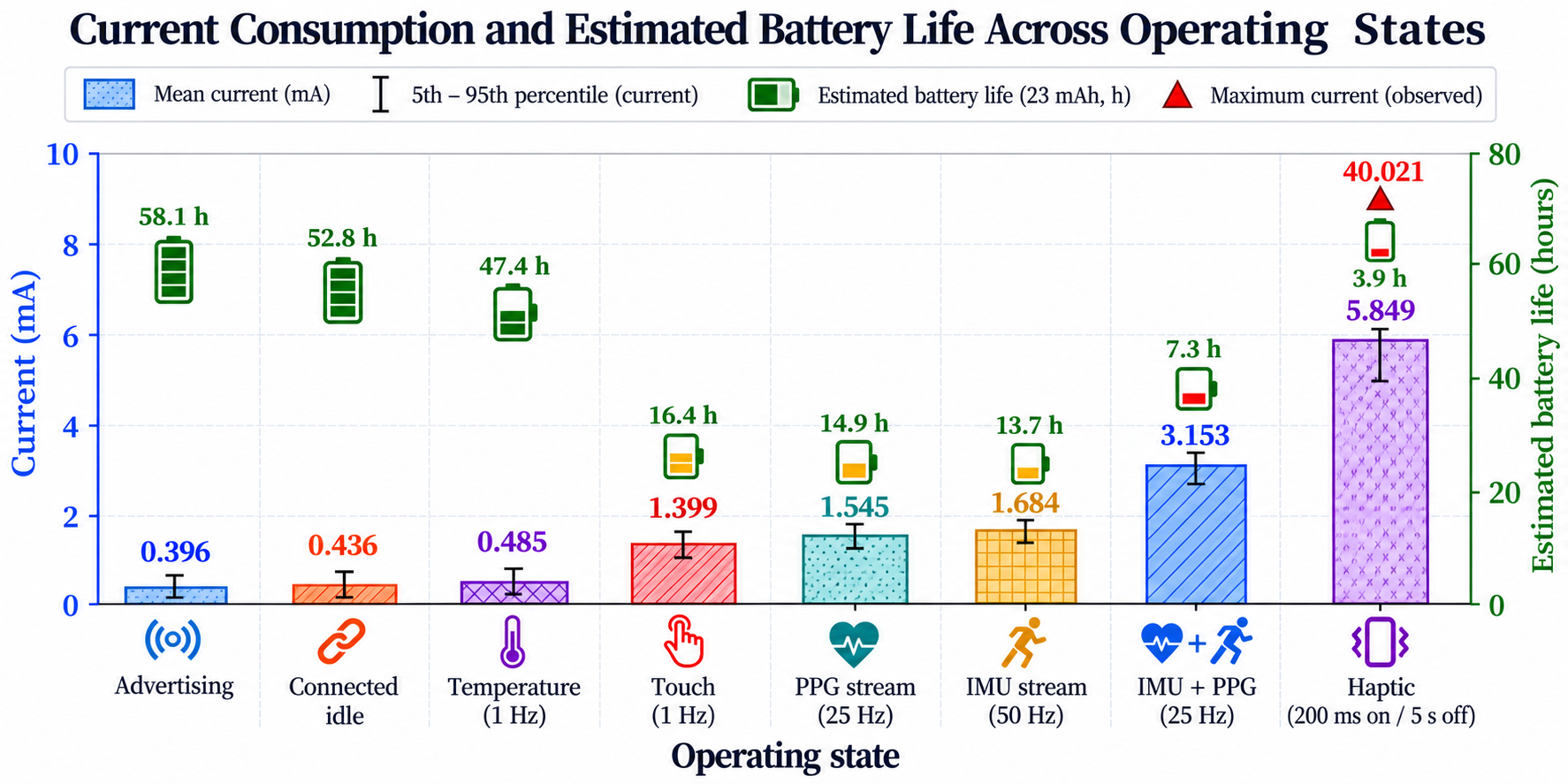}
    \caption{Measured DC input current across \platformname operating states. Battery-life values are projections and not measured discharge lifetimes.}
    \label{fig:current_operating_points}
    \vspace{-5 mm}
\end{figure}

We measured DC input current with a Keysight 34465A high-accuracy digital multimeter (DMM) under eight firmware workloads. Except for the haptic workload, the meter used autoranging and one power-line cycle integration; captures lasted \qtyrange{118.6}{176.6}{\second} at \qtyrange{16.3}{17.2}{\hertz}. The haptic capture used a fixed \qty{0.1}{\ampere} range and ten power-line cycle integration for \qty{123.4}{\second} at \qty{2.4}{\hertz}. Figure~\ref{fig:current_operating_points} reports the operating-state sample means and standard deviations. The projected battery life is calculated as $t_{\mathrm{est}}=\qty{23}{\milli\ampere\hour}/\bar{I}$. It assumes that the full rated capacity is usable under an unchanged workload and is therefore not a measured discharge lifetime.

As shown in Fig.~\ref{fig:current_operating_points}, advertising and connected-idle operation establish a sub-\qty{0.5}{\milli\ampere} baseline, while continuous sensing increases consumption depending on the active sensors, reaching \qty{3.1}{\milli\ampere} for simultaneous \gls{imu} and \gls{ppg} streaming. Low-rate body temperature sensing remains close to the idle baseline at \qty{0.48}{\milli\ampere}. Periodic haptic feedback is the most power-intensive workload; it averages \qty{5.8}{\milli\ampere} and produces short high-current periods during actuator activation, reaching a maximum observed sample of \qty{40.02}{\milli\ampere}. Overall, the idealized battery-life projections span \qtyrange{7.3}{58.1}{\hour} across the non-haptic workloads and reach \qty{3.9}{\hour} for the periodic haptic workload, although haptic consumption depends strongly on the load and its activation duty cycle.

\subsection{Streaming Completeness and Receiver Timing}

\begin{figure}[!t]
    \centering
    \includegraphics[width=\linewidth]{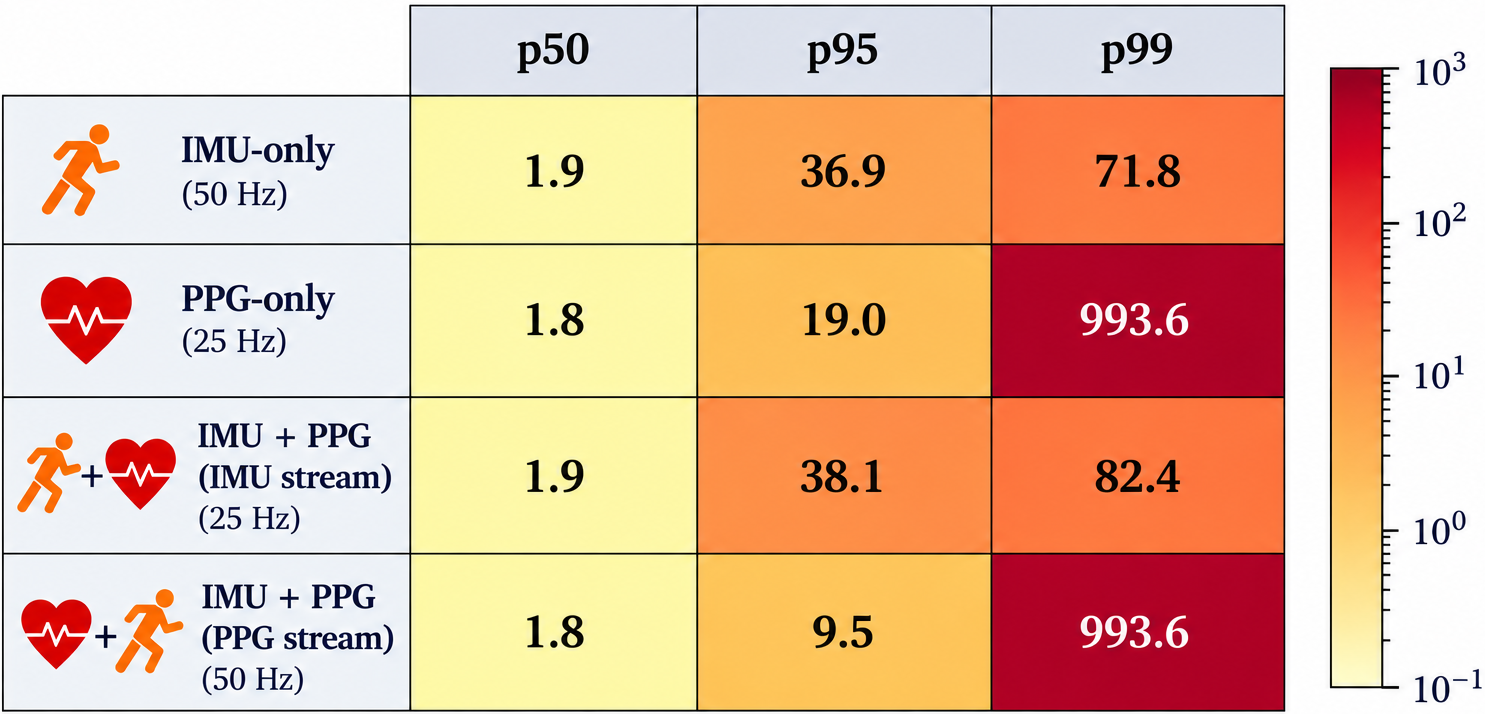}

    \caption{Receiver callback inter-arrival-time quantiles across the evaluated streaming configurations. Each cell reports the channel-mean interval in milliseconds; darker fill denotes a longer interval.}
    \label{fig:streaming_timing}
    \vspace{-4 mm}
\end{figure}

We evaluated end-to-end streaming in three \qty{10}{\minute} trials, i.e., \gls{imu}-only, \gls{ppg}-only, and simultaneous \gls{imu}+\gls{ppg}. All measured streams achieved 100\% data completeness, with matching counts from sensor production through reception and no observed internal drops, GATT-submission failures, or BLE delivery loss. Fig.~\ref{fig:streaming_timing} compares receiver callback inter-arrival times. The channel-mean $p_{50}$ values range from 1.83 to 1.92~ms; channel-mean $p_{99}$ values are 68.08~ms (\gls{imu}-only), 81.09~ms (\gls{imu} with \gls{ppg}), 990.42~ms (\gls{ppg}-only), and 992.81~ms (\gls{ppg} with \gls{imu}). The \gls{ppg} tail is consistent with burst-like application callback delivery.

\begin{figure}[!t]
        \centering
        \includegraphics[width=1\linewidth]{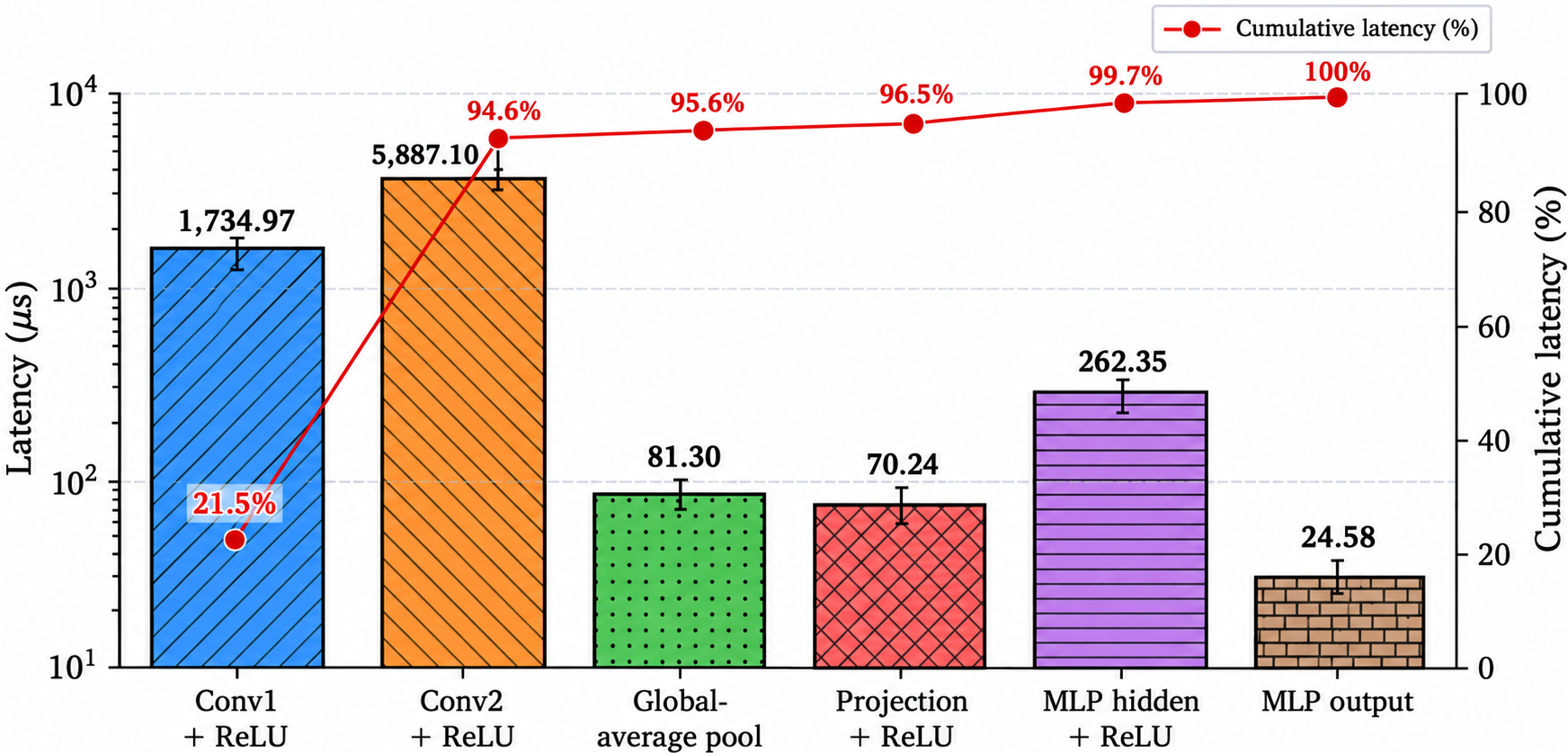}
        \label{fig:placeholder}
        \caption{Mean per-stage execution latency of the representative \gls{ai} model across 3,000 measured executions using fixed-seed synthetic inputs and weights.}
    \label{fig:ai_compute_stages}
    \vspace{-5 mm}
\end{figure}

\begin{figure}[!t]
    \centering
    \includegraphics[width=\linewidth]{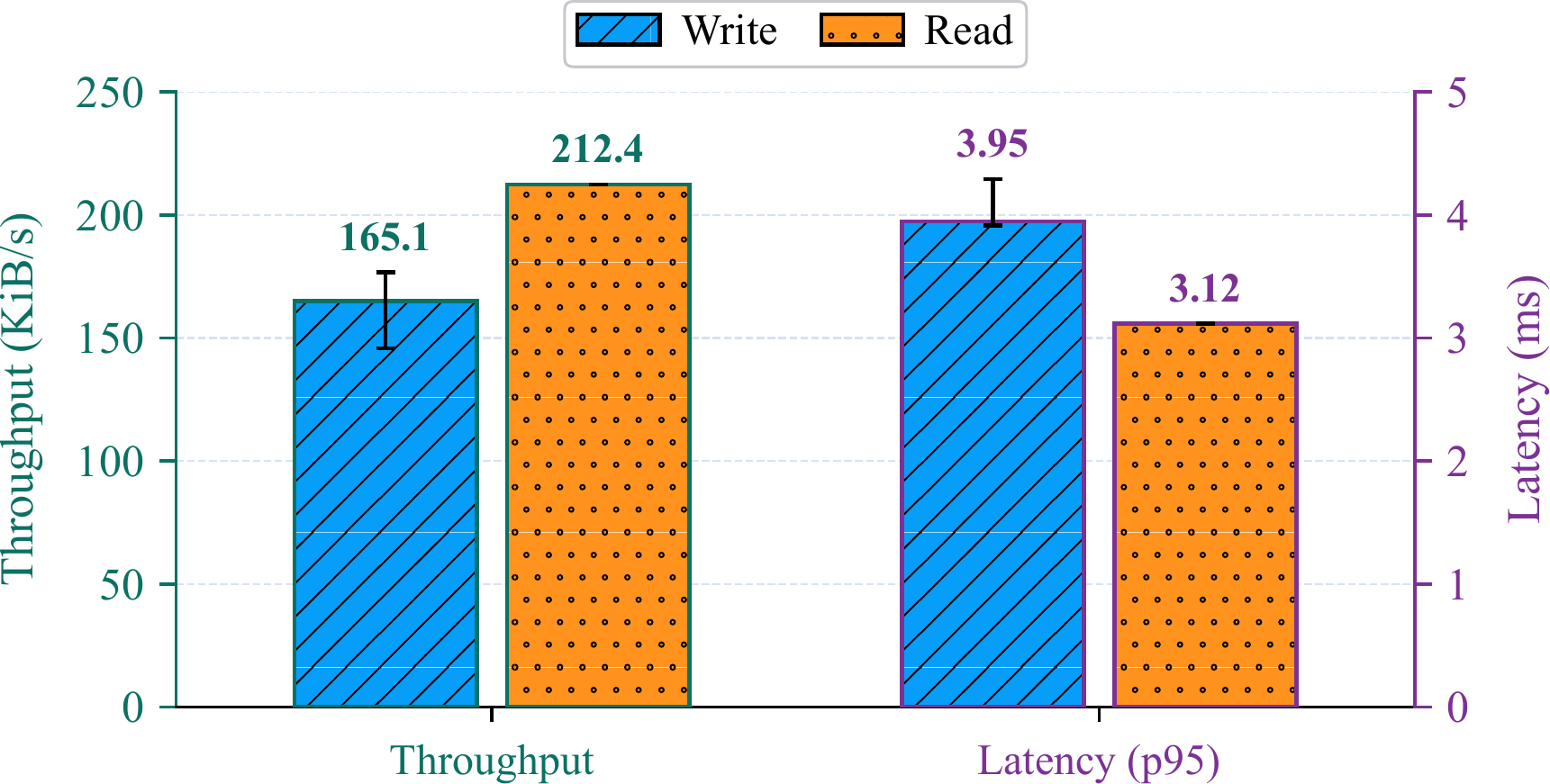}
    \caption{Local-storage performance showing sequential read/write throughput and application-call
    $p_{95}$ latency.}
    \label{fig:storage_performance}
    \vspace{-5 mm}
\end{figure}

\subsection{Local Storage}
\platformname integrates a \qty{4}{\mebi\byte} M95P32 nonvolatile memory for local data recording. To characterize the available storage performance, firmware on the nRF54L15 sequentially wrote and read a \qty{128}{\kibi\byte} LittleFS file using \qty{512}{\byte} block size and maximum \gls{spi} frequency of 8MHz, with ten repetitions. As shown in Fig.~\ref{fig:storage_performance}, the throughput is approximately 165 KiB/s for writes and 212 KiB/s for reads. Moreover, the latency for reads is 3.12 ms and for writes is 3.95 ms. All runs passed file-size, byte-count, and CRC32 checks, validating local storage operation through the firmware and filesystem stack.

\subsection{Representative \gls{ai} Model Compute}

To characterize the compute capacity available for on-device AI, we benchmarked a representative lightweight neural network on the nRF54L15 using a fixed-seed synthetic $3\times128$ input and FP32 weights. The model consists of two Conv1D/ReLU stages, global-average pooling, a 64-dimensional projection, and a $64\rightarrow64\rightarrow6$ MLP. Each of three boots used 100 warm-up executions followed by 1,000 measured executions. As shown in Fig.~\ref{fig:ai_compute_stages}, the complete model executes in \qty{8.0}{\milli\second} on average, supporting up to \num{124.1} inference windows/s. The two convolution stages dominate execution, requiring \qty{1.7}{\milli\second} and \qty{5.8}{\milli\second} and together accounting for \qty{94.6}{\percent} of the total latency. For a \qty{100}{\hertz} sensor stream with 128-sample windows and 50\% overlap, only \num{1.5} inference windows/s are generated, corresponding to a \qty{1.2}{\percent} compute duty cycle and leaving substantial headroom for additional on-device processing. All outputs matched the Python FP32 reference within the predefined $10^{-4}$ tolerance. Since synthetic inputs and weights are used, these results characterize on-device compute capability rather than application-level \gls{ai} accuracy.

\section{An End-to-End Case Study on \platformname: Resting Heart-rate Estimation}
\label{sec:case_studies}
\begin{figure}
    \centering
    \includegraphics[width=0.9\linewidth]{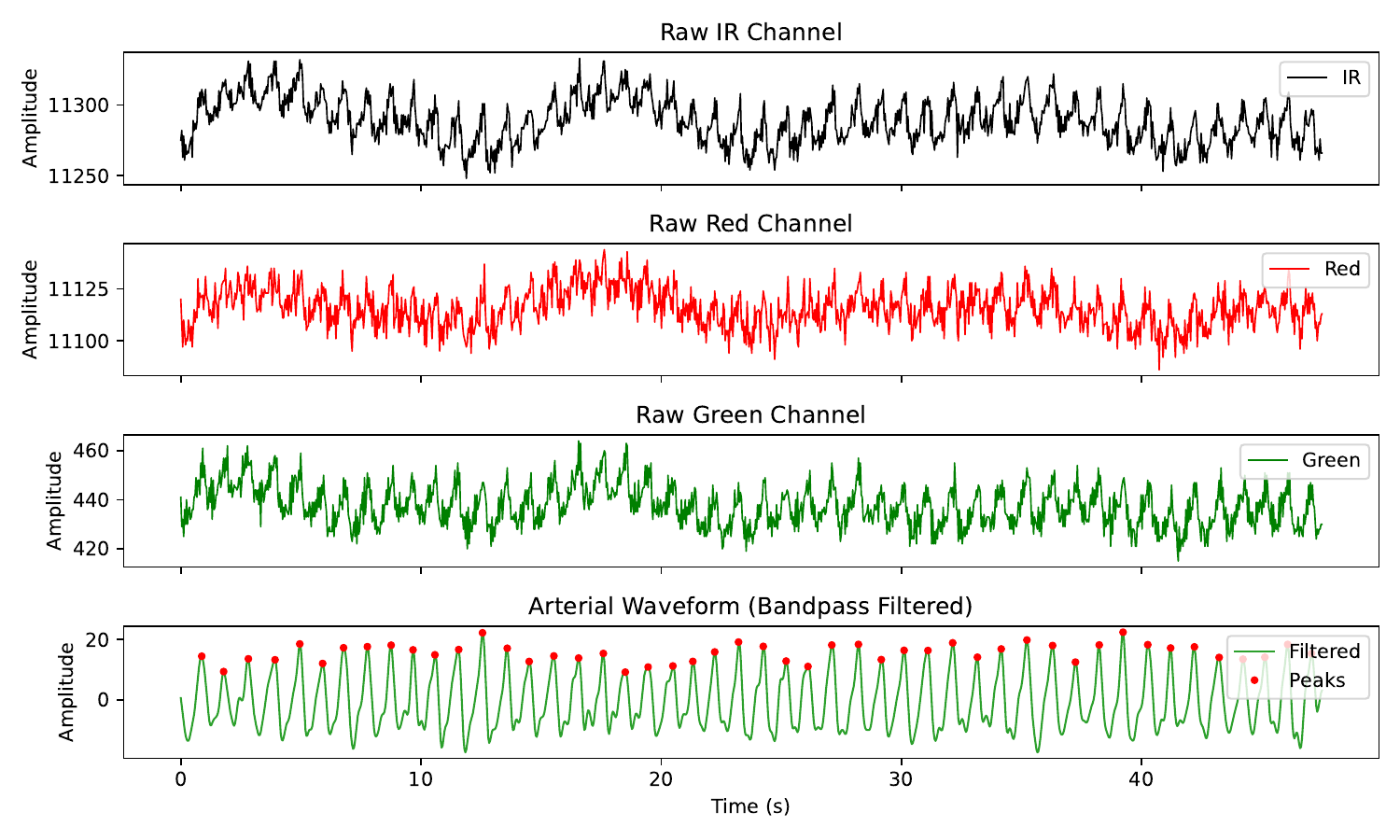}
    \caption{Raw \gls{ppg} signals (\gls{ir}, red, and green) acquired using the \platformname ring and the resulting bandpass-filtered arterial waveform with detected peaks.}
    \label{fig:ppg_waveform_example}
    \vspace{-5 mm}
\end{figure}

We demonstrate the end-to-end use of \platformname with real sensor data through a physiological case study: arterial pulse waveform analysis and resting heart-rate estimation. For this case study, \platformname was used in its ring form factor, with the \gls{ppg} daughter board acquiring the optical signals and the platform's firmware and software stack collecting raw sensor data and performing the necessary estimation processing. The results are compared with external reference devices. For this experiments, the participant gave written informed consent, and the protocol was approved by the relevant institutional ethics committee. The \gls{ppg} sensor daughter board collect raw data including IR, red, and green optical channels. Fig.~\ref{fig:ppg_waveform_example} shows the acquired multi-channel optical signals and the resulting bandpass-filtered arterial waveform with detected systolic peaks. 

\begin{table}[!t]
\caption{Resting heart-rate comparison with the Omron reference.}
\label{tab:hr_comparison}
\centering
\begin{tabular}{lccc}
\toprule
Device & MAE (bpm) & RMSE (bpm) & Correlation \\
\midrule
\platformname (Ours) & 1.8 & 2.3 & 0.88 \\
Shimmer & 1.8 & 2.6 & 0.84 \\
Oura Ring & 2.2 & 2.6 & 0.92 \\
\bottomrule
\end{tabular}
\vspace{-6 mm}
\end{table}

\begin{table*}[!t]
\centering
\caption{Comparison of \platformname with existing wearable research platforms and consumer devices.}
\label{tab:platform_comparison}
\scriptsize
\begin{tabular}{lccccccccccc}
\toprule
\textbf{Platform} & \textbf{Open} & \textbf{Raw} & \textbf{\gls{imu}} & \textbf{\gls{ppg}} & \textbf{Other} & \textbf{Custom} & \textbf{Edge} & \textbf{Modular} & \textbf{Form-factor} & \textbf{Actuators} & \textbf{MIT}\\
 & \textbf{Source} & \textbf{Data} &  &  & \textbf{sensors} & \textbf{Sensors} & \textbf{AI} &  & \textbf{Agnostic} & & \textbf{License} \\
\midrule

\textbf{\platformname (Ours)} 
& \checkmark & \checkmark & \checkmark & \checkmark & Temp/Touch & \checkmark 
& \checkmark & \checkmark & \checkmark & LED/Haptic & \checkmark \\

\midrule
\multicolumn{12}{l}{\textit{Research platforms}}\\
\midrule

BioGAP~\cite{frey2023biogap} 
& \checkmark & \checkmark  & \checkmark  & \checkmark & Temp/EEG 
& \checkmark & Limited & \checkmark & $\times$ (fixed forms) 
& $\times$ & \checkmark \\

BioGAP-Ultra~\cite{frey2026biogap}
& \checkmark & \checkmark & \checkmark & \checkmark & Temp/EEG/Mic. 
& \checkmark & Limited & \checkmark & $\times$ (fixed forms) 
& $\times$ & \checkmark \\

OpenEarable~\cite{roddiger2022openearable}
& \checkmark & \checkmark & \checkmark & -\checkmark & Temp/Pres./Mic. 
& \checkmark & \checkmark & $\times$ & $\times$ (earable) 
& LED & \checkmark \\

OpenEarable 2~\cite{roddiger2025openearable}
& \checkmark & \checkmark & \checkmark & \checkmark & Temp/Pres./Mic. 
& \checkmark & \checkmark & $\times$ & $\times$ (earable) 
& LED & \checkmark \\

OmniRing~\cite{zhou2023one}
& \checkmark & \checkmark & \checkmark & \checkmark & $\times$ 
& \checkmark & Limited & $\times$ & $\times$ (ring only) 
& LED & $\times$ \\

Lumos~\cite{watson2023lumos}
& \checkmark & \checkmark & $\times$ & $\times$ & Spectroscopy 
& $\times$ & Limited & $\times$ & Partial (2 forms) 
& $\times$ & \checkmark \\

OpenSenseRT~\cite{slade2021open}
& \checkmark & \checkmark & \checkmark & $\times$ & $\times$ 
& $\times$ & Limited & $\times$ & $\times$ (body \gls{imu}s) 
& LED & $\times$ \\

Empatica~\cite{mccarthy2016validation}
& $\times$ & \checkmark & \checkmark & \checkmark & Temp/EDA 
& $\times$ & $\times$ & $\times$ & Partial (2 forms) 
& LED/Haptic & $\times$ \\

Shimmer~\cite{burns2010shimmer}
& $\times$ & \checkmark & \checkmark & \checkmark & ECG/EMG 
& $\times$ & $\times$ & $\times$ & Fixed forms 
& $\times$ & $\times$ \\

\midrule
\multicolumn{12}{l}{\textit{Consumer devices}}\\
\midrule

Apple Watch
& $\times$ & $\times$ & \checkmark & \checkmark & Temp/ECG 
& $\times$ & $\times$ & $\times$ & $\times$ (watch) 
& LED/Haptic & $\times$ \\

Oura Ring
& $\times$ & $\times$ & \checkmark & \checkmark & Temp 
& $\times$ & $\times$ & $\times$ & $\times$ (ring) 
& $\times$ & $\times$ \\

\bottomrule
\end{tabular}
\vspace{1mm}
\footnotesize
\vspace{1mm}
\textbf{Legend:} 
Open Source = hardware + firmware availability; Raw Data = synchronized low-level access; Custom Sensors = ability to integrate new modules; Edge \gls{ai} = on-device model support; Modular = interchangeable sensing/actuation; Form-factor Agnostic = not tied to one embodiment.
\vspace{-5 mm}
\end{table*}

We evaluate resting heart-rate estimation using \platformname against a Shimmer \gls{ppg} device and an Oura Ring, with an Omron cuff-based blood-pressure monitor used as a reference. All devices were worn simultaneously during rest, and heart-rate measurements were synchronized by measurement interval. Performance is evaluated using mean absolute error (MAE), root mean squared error (RMSE), and Pearson correlation with respect to the Omron reference. The results in Table~\ref{tab:hr_comparison} show that \platformname provides heart-rate estimates comparable to the two external wearable devices, with the same MAE as Shimmer, lower RMSE than both comparison devices, and a correlation between those observed for Shimmer and Oura. Overall, this case study demonstrates the end-to-end use of \platformname from raw \gls{ppg} acquisition and on-device signal processing to resting heart-rate estimation, while retaining access to the underlying raw sensor data.

\section{Related Work}
\label{sec:related_work}
Wearable platforms typically trade measurement quality for experimental flexibility. Application-specific systems such as OpenSenseRT, BioGAP, BioGAP-Ultra, Lumos, and the Open Speech Platform provide strong sensing for motion, biosignals, spectroscopy, or hearing research, but their hardware and placement are task-specific~\cite{slade2021open,frey2024wearable,frey2026biogap,watson2023lumos,pisha2019wearable}. Open devices such as OmniRing and OpenEarable improve reproducibility, yet still bind the sensing stack to ring or ear embodiments~\cite{zhou2023one,roddiger2022openearable,roddiger2025openearable}. Broader surveys identify interoperability, energy efficiency, privacy, data management, and clinical/cloud integration as persistent barriers for wearable \gls{ai}~\cite{ometov2021survey,ghadi2025integration}. \platformname targets the gap between these categories by exposing modularity at the daughter-board boundary while keeping the main board, firmware abstractions, raw data stream, and feedback interfaces stable across embodiments.

Table~\ref{tab:platform_comparison} compares data access, sensing and actuation, extensibility, and licensing. The systems cover a range of sensing domains, while \platformname combines interchangeable daughter boards, reusable electronics across form factors, raw data access, and permissive reuse to support experiments that vary sensor configuration and placement.

\section{Discussion and Conclusion}
\label{sec:discussion_conclusion}
\platformname combines reusable electronics with configurable sensing and a common data interface. The shared power and signal connections in Fig.~\ref{fig:hardware_architecture} allow daughter-board changes while retaining the main board. The driver, device-manager, and GATT layers in Figure~\ref{fig:firmware_architecture} carry those configurations through to timestamped data and control interfaces. Together, these boundaries support experiments that vary sensing, placement, feedback, and processing location. The implementation repositories are available through the \platformname GitHub page (link omitted for double-blind review).

The characterization identifies capabilities and bottlenecks, while the battery-life annotations are capacity-to-mean-current projections rather than discharge-test results. In the three 10-minute streaming trials, all four measured boundary counts matched for every channel, with no observed internal drops or BLE delivery loss. Receiver inter-arrival quantiles describe application callbacks, not RF latency or range. The storage result is sequential and CRC-validated, not an endurance claim; the fixed-seed synthetic candidate model measures compute headroom, not application accuracy. An end-to-end case study demonstrates raw \gls{ppg} waveform access and competitive resting heart-rate estimates. Future studies should assess motion, skin tone, alternate placements, attachment stability, comfort, robustness, and battery life, with ECG or other appropriate references for each target quantity.

\section*{Acknowledgment}

OpenAI ChatGPT was used only for language editing and grammar improvement.
All content was reviewed and verified by the authors; \gls{ai} was not used for
system development, experiments, or result generation.

\IEEEtriggeratref{20}
\IEEEtriggercmd{\balance}
\bibliographystyle{IEEEtran}
\bibliography{references}

\end{document}